# Conformal Structure of Quantum Wave Mechanics

by Richard James Petti, ORCID: 0000-0002-9066-7751
146 Gray Street, Arlington MA 02476 USA
email: rjpetti@gmail.com, rjpetti@alum.mit.edu

July 16, 2022

Abstract

This work interprets the quantum terms in a Lagrangian, and consequently of the wave equation and momentum tensor, in terms of a modified spacetime metric. Part I interprets the quantum terms in the Lagrangian of a Klein–Gordon field as scalar curvature of conformal dilation covector nm that is proportional to ℏ times the gradient of wave amplitude R. Part II replaces conformal dilation with a conformal factor ρ that defines a modified spacetime metric g' = exp(ρ) g, where g is the gravitational metric. Quantum terms appear only in metric g' and its metric connection coefficients. Metric g' preserves lengths and angles in classical physics and in the domain of the quantum field itself. g' combines concepts of quantum theory and spacetime geometry in one structure. The conformal factor can be interpreted as the limit of a distribution of inclusions and voids in a lattice that cause the metric to bulge or contract. All components of all free quantum fields satisfy the Klein-Gordon equation, so this interpretation extends to all quantum fields. Measurement operations, and elements of quantum field theory are not considered.

## Contents

# 1   Introduction

In the-twentieth century, three seminal ideas came to dominate fundamental physics: geometry of parallel translation and curvature in gravitational theory, quantum theory, and unitary gauge theory. Quantum theory has three parts: quantum wave mechanics, measurement processes, and minimum action and discrete particles with energy given by Planck's formula $E = h\, f$, where $h$ is Planck's constant, and $f$ = frequency. Physicists have explored how conformal symmetry might play a role in quantum theory, despite its violation of metricity.[1]

This work defines two technical terms that are employed throughout.

"Complex polar form" means a wave function expressed using complex numbers in polar form, with functions $R$ = amplitude and $S$ = phase, as in equation (1). Spacetime coordinates are always Minkowski coordinates to simplify computation of results. Most results apply in curved spacetimes by replacing coordinate derivates with covariant derivatives..

(1) $$\psi(x) = R(x)\, \exp(-i\, S(x)) \qquad \text{(Apx A, d6)}$$

"Quantum terms" are terms in an expression that are proportional to $\hbar$ or $\hbar^2$ in a real variable polar representation using real variables $R$ = amplitude and $p_\mu$ = momentum, as in equation (2). The second equation in (2) assumes $S = p_\mu x^\mu / \hbar$.

(2) $$\psi(x) = R(x)\, \exp(-i\, p_\mu x^\mu) \qquad \text{where} \qquad p_\mu = \hbar\, S_{,\mu} \qquad \text{(Apx A, d7)}$$

We could almost define quantum terms as terms proportional to $\hbar$ or $\hbar^2$, except that a factor of $1/\hbar$ is contained inside $S$.

The main results in this work are achieved in two stages.

Part One establishes that the quantum terms in the Lagrangian of a Klein–Gordon field equals the scalar curvature of conformal dilation nm, plus a divergence that can be ignored in the action integral. nm is proportional to $\hbar$ times the gradient of the wave amplitude $R$. The key results are:

- The quantum terms occur only in the conformal dilation in the connection.
- The spacetime metric is conserved in classical physics, despite the use of conformal dilations.
- Conformal dilation can be interpreted geometrically as the limit of a continuum distribution of inclusions and voids in a lattice that causes the metric to bulge or contract.
- All components of all free quantum fields satisfy the Klein-Gordon equation, so this interpretation extends to all quantum fields.

Part One has these computational steps.

a) Express the KG Lagrangian using a wave function in complex polar form. See section 2 "The Klein-Gordon field."
b) Define the conformal dilation covector $nm_\mu$ as proportional to $\hbar\, R_{,\mu}$. This is the only place where $\hbar$ enters the theory when expressed in terms of $R$ and $\mathbf{p}$. See section 3.1 "The nonmetricity covector."
c) Compute the scalar curvature of the conformal dilation. See section 3.2 "Scalar curvature due to conformal dilation."
d) Verify that the scalar curvature of the conformal dilations is identical to the quantum term $\hbar^2\, R_{,\mu}\, R_{,\nu}\, g^{\mu\nu}$ plus a divergence.

Part Two replaces conformal dilation with a conformal factor $\rho$ that defines a modified spacetime metric $g' = \exp(\rho)\, g$, where $g$ is the gravitational metric. The key results are:

- The modified spacetime metric $g'$ is conserved by parallel translation in ordinary spacetime and in the domain of quantum waves. The modified metric removes the conformal dilation from the equations.
- Where there are no quantum waves, the metric $g'$ is identical to $g$.
- The quantum terms occur in the metric $g'$, not in the Lagrangian of the KG field.

- The field equations for this metric have the same solutions as the standard field equations for gravitation and a quantum wave function.

This work does not address other aspects of quantum theory: measurement, and discreteness of action, energy and particle number. This work does not present alternatives to standard computational methods of quantum mechanics that use complex wave functions, Hermitian operators, eigenvalues, eigenvectors, and projection operators.

Appendices: references to equations in Appendix A are referred to as "(Apx A, 'linenumber')"

- Appendix A: Computer algebra script for conformal dilation of the Klein-Gordon Lagrangian
- Appendix B: Affine connections with metric and conformal dilation

## 2  The Klein-Gordon field

### 2.1  Why use the Klein-Gordon field?

The main result in this work is that the quantum mechanical terms (those proportional to Planck's reduced constant ℏ) in the KG Lagrangian can be interpreted as the scalar curvature due to conformal dilation in the connection of spacetime. This result enables applying the geometric interpretation of conformal dilation to the quantum terms. The KG field enables us to derive the main results while avoiding dealing with spinors and z-spin.

In quantum field theory, the KG field cannot represent anything except a structureless particle with spin zero, of which we know only one: the Higgs boson. However, every field component of every free quantum field theory satisfies the KG equation, so the results can be extended to all quantum field theories.

### 2.2  Klein–Gordon Lagrangian, wave equation, and momentum tensor

We focus on the KG field in special relativity to derive the desired results with minimum of computational complexity. Let $x^\mu$ be coordinates and $g$ a metric tensor on a spacetime manifold $\Xi$ of dimension $dim$. The complex polar form of quantum wave equations dates back at least to the pilot wave model of quantum mechanics by David Bohm [2],[3].

We first present results in standard form to facilitate comparison with complex polar form.

The Lagrangian of the KG field in conventional form is:

(3) $$L = \frac{-g^{\alpha\beta}(-i\hbar\,\partial_\beta\psi)^\dagger(-i\hbar\,\partial_\alpha\psi) + m^2\psi^\dagger\psi}{m}$$ (Apx A, d2)

By varying the Lagrangian by the wave function $\psi$ to get the KG equation, we have

(4) $$(\hbar^2\,\psi_{,\alpha,\beta}\,g^{\alpha\beta} + m^2\,\psi)/m = 0$$ (Apx A, d3)

By varying the action $\int L\,\sqrt{(|\det(g)|)}\,d\mathbf{x}$ by metric $g_{\mu\nu}$ and divide by $\sqrt{(|\det(g)|}$ to get the momentum tensor. We get the special relativitist momentum tensor by varying the metric, then setting the metric equal to the Minkowski metric.

(5) $$T^{\mu\nu} = \frac{[\psi^\dagger m^2\psi - \hbar^2(\psi^\dagger_{,\alpha}\psi_{,\beta}\,g^{\alpha\beta})]\,g^{\mu\nu}}{2m} + \frac{\hbar^2(\psi^\dagger_{,\alpha}\psi_{,\beta}\,g^{\alpha\mu}\,g^{\beta\nu})}{m}$$ (Apx A, d5)

Express $\psi$ in complex polar form with amplitude $R(\mathbf{x})$ and phase $S(\mathbf{x})$. The Lagrangian (3) can be expressed in real variables $R$ and $\mathbf{p}$.

(6) $$L = \frac{R^2(m^2 - p^2) - \hbar^2\,g^{\nu\mu}\,R_{,\nu}\,R_{,\mu}}{m}$$ (Apx A, d8)

where momentum vector $\mathbf{p}$ is defined in equation (2). The KG Lagrangian without its quantum terms is

(7) $$L' = R^2(m^2 - p^2)/m$$

Multiply the KG equation by $\psi^\dagger$ and separate the real and imaginary parts. The real part of the KG equation is the relativistic mass-momentum relationship including the quantum term.

(8) $$\frac{R^2 (m^2 - p^2) + \hbar^2 R R_{,\alpha,\beta} g^{\alpha\beta}}{m} = 0 \qquad \text{(Apx A, d10)}$$

The imaginary part of the equation is the relativistic conservation equation for the momentum vector.

(9) $$\frac{\hbar (R^2 p^\lambda)_{,\lambda}}{m} = 0 \qquad \text{(Apx A, d11)}$$

All polar forms can be expressed in terms of squared amplitude $n \equiv R^2$ and 4-momentum $p$ using the substitutions

(10) $\qquad R_{,\lambda} \to \tfrac{1}{2} n_{,\lambda}/\sqrt{n} \qquad$ and its inverse $\qquad n_{,\lambda} \to 2 R R_{,\lambda}$

### 2.3 Minimal electromagnetic coupling

A real–variable formulation of minimal electromagnetic coupling can be added to the Lagrangians and field equations of quantum wave fields by replacing equation (2) with the kinetic momentum

(11) $$p_\mu = \hbar S_{,\mu} - q A_\mu$$

where $q$ = charge of the field and $A_\mu$ = electromagnetic 4-potential.[4]

Defining $p$ as in equation (11) solves two problems.

a) $p_\mu = \hbar S_{,\mu} - q A_\mu$ is invariant under U(1) gauge transformations. The effects of change in U(1) gauge on these two terms are equal in magnitude and opposite in sign.
  - $-\partial_\mu S$ is the rate of change of phase of a wave field without electromagnetism under translation in the $x^\mu$ direction.
  - $q A_\mu/\hbar$ is the rate of change of phase of a wave field due to the electromagnetic field under translation in the $x^\mu$ direction.

b) The Lorentz force requires that the covector field $p$ has nonzero exterior derivative (curl).

The following computation illustrates why (b) is true. Choose coordinates so that $x^0$ is in the direction of $p$, and $x^1$ is in the direction of the electric field E. The Lorentz force equation becomes (where $\tau$ = proper time):

(12) $$\partial_\tau \begin{bmatrix} p^0 \\ p^1 \end{bmatrix} = q/m \begin{bmatrix} F^0{}_1 p^1 \\ F^1{}_0 p^0 \end{bmatrix} = q/m \begin{bmatrix} 0 \\ F^1{}_0 p^0 \end{bmatrix}$$

The exterior derivative of $p$ becomes approximately

(13) $$\text{ext\_diff}(p)_{01} = \partial_0 p_1 - \partial_1 p_0 = q/m\, F^1{}_0\, p^0 = q/m\, E^1\, p^0$$

Recall that, in its rest frame, a particle does not interact with magnetic fields, so equation (13) is sufficient to determine the exterior derivative of $p$ in general spacetime coordinates.

(14) $$\text{ext\_diff}(p)_{\mu\nu} = \partial_\mu p_\nu - \partial_\nu p_\mu = q/m\, F_{\mu\nu}$$

The electromagnetic field adds a term to the Lagrangian[5]

(15) $$L_{EM} = - 1/4\, F_{\alpha\beta}\, F^{\alpha\beta}.$$

## 3 From quantum waves to conformal dilation

The objective of this section is to define conformal dilation, compute the scalar curvature due to conformal dilation, and show that the result equals the quantum term in the KG Lagrangian.

### 3.1 The nonmetricity covector

Define the conformal dilation covector (nm for "nonmetricity") as proportional to $\hbar$ and the gradient of R.

(16) $$nm_\mu = \frac{2 \sqrt{(2\, K_{grav})}\, \hbar\, R_{,\mu}}{\sqrt{((\dim - 2)(\dim - 1)\, m)}} \qquad \text{(Apx A, d27)}$$

The other constants are chosen as follows.

a) The covector nm includes a factor $\sqrt{(2\, K_{grav})}$ because the field equations of general relativity are $G_{\mu\nu} / (2\, K_{grav}) = momentum_{\mu\nu}$, where $K_{grav} = (8\pi G / c^4)$. Although we compute only in special relativity, this constant gives nonmetricity the correct units to fit into general relativity.

b) Factors containing dim enable nm to replicate the quantum terms in the KG equation in spacetime of any dimension.

### 3.2 Scalar curvature due to conformal dilation

In special relativity, the scalar curvature due to conformal dilation becomes

(17) $$\text{Scalar curvature} = \frac{-2\, K_{grav}\, \hbar^2\, R_{,\mu}\, R_{,\nu}\, g^{\mu\nu} + \text{constant}\, R_{,\mu,\nu}\, g^{\mu\nu}}{m} \qquad \text{(Apx A, d28)}$$

The second term on the right side of equation (17) equals $R_{,\mu}{}^{;\mu}$ (with a covariant derivative denoted ";"), which is a covariant divergence. In an action integral, this becomes a boundary term which drops out. Therefore, in a spacetime of dim ≥ 2, equation (17) becomes:

(18) $$\text{Scalar curvature}/(2\, K_{grav}) = -\, R_{,\mu}\, R_{,\nu}\, g^{\mu\nu}\, \hbar^2/m \qquad \text{(Apx A, d30)}$$

This equals the quantum term in the KG Lagrangian when the KG field is written in complex polar form $\psi(x) = R(x)\, \exp(-i\, S(x))$.

### 3.3 Nonmetricity

Any extension of affine symmetry beyond Einstein–Cartan theory (which equates affine torsion with intrinsic angular momentum) must use symmetries that do not preserve the metric: conformal dilations, projective (special) conformal transformations, and/or non-conformal nonmetricity. This work does not use affine torsion or Einstein–Cartan theory; however, the expression for the coefficients of the full connection with torsion and conformal dilation is given in Appendix B equation (B-5).

The following construction illustrates that conformal dilation (16) does not violate metricity at the scale of classical spacetime physics.

- Consider the effect of conformal dilation along a radial path that passes through the center of a Gaussian wave packet starting and ending on opposite sides of the wave packet. The metric distortion is proportional to

(19) $$\hbar \int (dR/d\rho)\, d\rho = R(\rho_2) - R(\rho_1)$$

where R is the amplitude of the wave packet, $\rho$ is a path parameter that is zero at the center of the wave packet, and $\rho_1$ and $\rho_2$ are the values of the curve parameter at the ends of the path. If the lengths of the two halves of the path are equal, then there is zero metric distortion outside of the wave packet.

- Consider parallel translation along any part of a circular path around the center of a Gaussian wave packet. The direction of motion is orthogonal to the radial direction of the conformal dilation covector; so lengths of vectors do not change.

### 3.4 Geometric interpretation of conformal dilations

Conformal dilation means that the connection preserves the conformal structure metric $g_{\mu\nu}$, but not all of $g_{\mu\nu}$.

(20) $$g_{\mu\nu;\beta} = -\, nm_\beta\, g_{\mu\nu} \qquad \text{(Apx B, B-6)}$$

It can be interpreted as the continuum limit of a distribution of inclusions and voids in an affine lattice. Figure 1 is a schematic of an inclusion in a 2D lattice; the inclusion causes the cell where it is located to bulge and to have greater area than a normal cell. Figure 2 is a schematic of a void in a 2D lattice; the void causes the cell where it is located to shrink and to have less area than a normal cell.

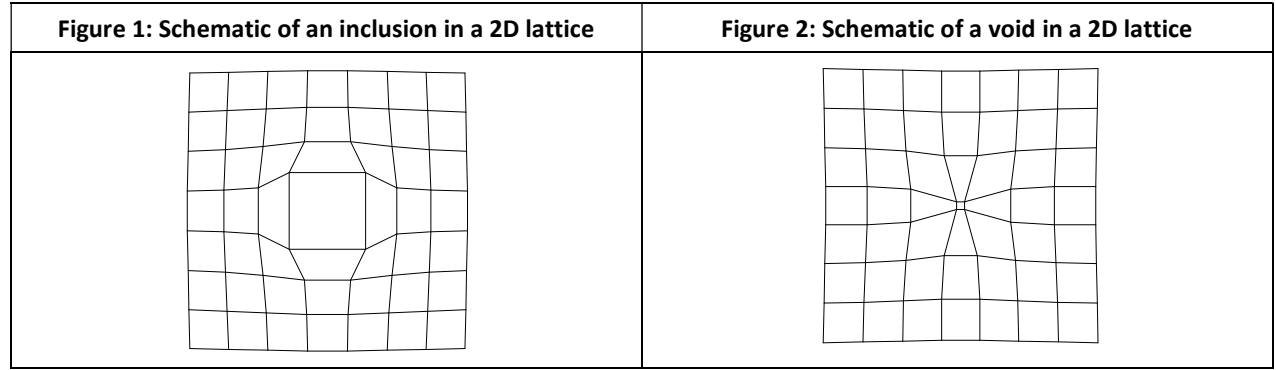

| Figure 1: Schematic of an inclusion in a 2D lattice | Figure 2: Schematic of a void in a 2D lattice |

See reference[6] for illustrations of dislocations, dislocations, and dispirations.

We normally consider metric affine geometry as intrinsic to spacetime. However, in these computations, conformal dilation is part of the geometry that is determined by the quantum matter fields.

## 4 From conformal dilation to modified spacetime metric

### 4.1 Conformal factor and modified spacetime metric

This theory can be restated with a modified spacetime metric without using conformal dilation. Equation (16) defines nm to be proportional to grad(R); R is a globally defined function; and spacetime is assumed to be simply connected (has no holes or loops). So nm is integrable to a scalar conformal factor, denoted by ρ.

(21) $$\rho = \frac{2\sqrt{2 K_{grav}}\ \hbar\ R}{\sqrt{(dim-2)(dim-1)}\ m}$$

The modified spacetime metric is

(22) $$g'_{\mu\nu} = \exp(\rho)\ g_{\mu\nu}$$

g' has these properties:

- $g'_{\mu\nu}$ is preserved by the connection.

(23) $$g'_{\mu\nu;\lambda} = (-nm_\lambda + \rho_{,\lambda})\exp(\rho)\ g_{\mu\nu} = (-nm_\lambda + \rho_{,\lambda})\ g'_{\mu\nu} = 0$$

- $g'_{\mu\nu}$ equals $g_{\mu\nu}$ everywhere in spacetime except in regions where a quantum field is present.
- The spacetime connection is the metric connection of g'. The Weyl term $CW_{\beta\mu\nu}$ defined in Appendix B equation (B-7) converts the metric connection of g to the metric connection of g'.

Even if spacetime inside the domain of a quantum field is not simply connected, and if nm generates the zero element of the first cohomology group of spacetime, then the function ρ and the metric g' exist, and a continuum limit of inclusions and voids accurately describes the effects of nm or ρ.

Like the formulation with conformal dilation, this metric formulation straddles the divide that separates quantum theory and spacetime geometry: the modified metric g' contains the conformal factor ρ that arises from quantum wave fields.

### 4.2 Several quantum fields and several quantum wave functions

This section clarifies that ways to combine different wave functions are the same as in standard quantum mechanics. However, in the conformal representation, the wave functions can have different affine connections. Let

(24)     $\psi 1 = R1\ \exp(-i\ S1)$     and     $\psi 2 = R2\ \exp(-i\ S2)$

Case 1, superposition: If ψ1 and ψ2 contribute to the same component of a quantum field, then the wave functions are superimposed.

(25) $$\psi 12 = R12\ \exp(-i\ S12)$$

(26) $$R12 = \sqrt{R12^2 - 2\cos(S2 - S1) + R2^2}$$

(27)     $S_{12} = \text{atan2}(-(R_1 \sin(S_1) + R_2 \sin(S_2)), R_1 \cos(S_1) + R_2 \cos(S_2))$

The wave functions are combined in one nonmetricity covector $nm_{12} = nm_1 + nm_2$, or in one conformal factor $\rho_{12} = \rho_1 + \rho_2$.

Electromagnetism serves as a model for interactions mediated by unitary gauge fields.

- The gauge potentials (connection coefficients) multiplied by the relevant charges are included in covariant derivatives in the Lagrangian for a charged field as in equation (11).
- The gauge field has its own term similar to equation (15) in the Lagrangian.

Case 2, tensor products: If $\psi_1$ and $\psi_2$ contribute to different components of the same quantum field or different quantum fields, then wave function are combined in a tensor product, and the Lagrangian is the sum of the Lagrangians for the separate fields, plus terms describing any interactions between the fields.

This situation is analogous to that in a multi-component crystal lattice. Consider a cubic lattice of sodium chloride, in which one sub-lattice populated by sodium ions, and another by chloride ions. If we apply an electric field at the resonant frequency of the two sub-lattices, they oscillate against one another. An electric field at the resonant frequency of one sub-lattice can induce oscillations in that sub-lattice. However on a larger scale, the material has one metric affine structure.

### 4.3    Field equations

This section presents two simple cases of Lagrangian and field equations in a format that would fits with general relativity.

For the first and simpler case, the modified spacetime metric is given in equations (21) and (22).

The Lagrangian is

(28)     $RS' + K_{grav}\, KG\_Lagr'$

where

- $RS'$ = scalar curvature of the torsion-free metric connection of the modified metric $g'$
- $KG\_Lagr'$ = KG Lagrangian without the quantum terms as shown in equation (7).

Varying the Lagrangian (28) by modified metric $g'$ and wave amplitude $R$ yields the field equations of gravitation and the KG field. The solutions of these field equations are identical to the solutions of the standard field equations of gravitation and the KG field.

Phase does not occur in this example because there is only one wave component. Below is a slightly more complicated example that includes phase functions, using a superposition of two quantum waves for fields of the same type.

The second case includes two quantum wave functions for a field of the same type. It is well-known that if we have two quantum wave fields, the same properties as listed for the first model with one wave function holds for this model, and the phase functions occur in the modified metric $g'$ and its connection coefficients.

In both cases, the main difference from standard treatments is that terms proportional to $\hbar$ are present only in the conformal factor $\rho$ in equation (21) and therefore in the modified metric (22), and of course in the connection coefficients that contain $g'$.

It is not clear whether this approach will contribute to computational methods in quantum theory. This approach does provide view of quantum mechanics that is compatible with the geometric methods of gravitational theory.

### 4.4    Symplectic structure

Symplectic geometry, along with its applications in Hamiltonian and Lagrangian mechanics, is at this time the best overall geometric treatment of phase of waves in classical and quantum physics. The interpretation of quantum waves as conformal dilations or modified spacetime metric links the quantum terms to the metric geometry of spacetime, which symplectic structure does not address. We include a brief description of symplectic structure for comparison.

After differentiable manifolds, the most basic geometric concept in mechanics is symplectic structure. Every smooth manifold $M$ has on its cotangent bundle $T^*M$ a canonical symplectic structure. ("Canonical" means the definition does not depend on human choices; this is a rare property among mathematical structures in physics.) For a smooth manifold M, the canonical symplectic 1–form is:

(29) $$\alpha = p_j \, dq^j$$

where $\{q^j\}$ are Cartesian coordinates on M and $\{p_j\}$ are the coordinates on T*M such that $p_j(\partial/\partial q^k) = \delta^k_j$.

Each momentum component $p_j$ is the rate of change of phase when moving in direction $\partial/\partial q^j$. So α is the rate of change of phase when moving along a path $(q(\tau), p(\tau))$ in T*M (phase space). The basic variational computation of symplectic mechanics finds path(s) $(q(\tau), p(\tau))$ in T*M such that nearby paths have nearly identical rates of phase change, so the nearby paths constructively interfere. The variational equation is:

(30) $$\delta \int p_j \, dq^j \, / \, \delta q(\tau) = 0 \qquad \tau = \text{proper time}$$

Equations (30) is the "stationary action principle" (or incorrectly, the principle of least action), because a solution path is a critical point of the action integral. This equation yields the basic equations of classical mechanics – even where no phase is evident – because quantum mechanics underlies all of classical physics.

Equation (30) yields Hamilton's equations.

(31) $$i \hbar \, \varphi^\bullet = - (\hbar^2/2m) \, \nabla^2 \varphi \qquad p^\bullet_k = - i \hbar \, \partial \varphi / \partial q^k \qquad H^\bullet = 0 \qquad \text{``}\bullet\text{''} = d/dt$$

When transformed by a Legendre transformation to the tangent bundle with paths $(q(\tau), q^\bullet(\tau))$, equation (30) yields Lagrange's equations.

(32) $$\frac{\partial L}{\partial \psi} - \nabla\left(\frac{\partial L}{\partial \nabla \psi}\right) - \frac{\partial}{\partial t}\left(\frac{\partial L}{\partial \psi^\bullet}\right) = 0$$

The canonical symplectic 2-form ω is the exterior derivative of α as a 1-form on the manifold T*M.

(33) $$\omega = d\alpha = dp_j \wedge dq^j$$

ω is the geometric version of the uncertainty principle of quantum mechanics in its covariant form in phase space.

(34) $$\omega(\Delta p, \Delta q) \geq \hbar/2$$

Inequality (34) is called the "uncertainty principle" because it describes the minimum uncertainty in simultaneous measurements of position and momentum. The left side of inequality is the symplectic area of a region in phase space defined by a pair of conjugate vectors; $\hbar/2$ is the minimal action that occurs in quantum theory. Therefore this inequality deserves the name "principle of least action," though this name is already incorrectly used for the "stationary action principle."

ω also defines a canonical complex structure $J_i^j$ on T*M, which is the complex number *i* in quantum mechanics.

(35) $$\omega_i^j \, \omega_j^k = - \text{Id}_i^k \qquad \text{that is,} \qquad J_i^j \, J_j^k = - \text{Id}_i^k$$

Symplectic structure is unique in the sense that it is the most comprehensive theory that applies from classical mechanics to quantum field theory. Symplectic geometry underlies most key structures of quantum theory – wave equations, definition of action, and complex structure.

This work focuses on interpretation of quantum waves in terms of conformal dilations in Part One, and a modified spacetime metric in Part Two. Symplectic structure makes no such link between wave mechanics and spacetime metric geometry; defining this link appears to likely the main contribution of this work.

The symplectic and conformal structures in quantum wave theory are in a sense complementary. The symplectic structure is defined by the gradient along a path of the phase of the complex wave function, and the conformal dilations are defined by the gradient of the amplitude of the complex wave function.

The symplectic forms α and ω can be interpreted as the connection form and the curvature form respectively of a connection on a circle bundle over phase space. From this point of view, the uncertainty principle (34) (which should be called the principle of least action) means that every measurable action is least one complete rotation in this circle bundle. The author conjectures that the uncertainty principle should be extended to a "principle of discrete action" in units of ℏ/2, so that any measurable action consists of an integer number of rotations around the circle bundle, and continuous geometry breaks down in phase space.

## 5    Further developments

Extend the construction from the KG equation to the Dirac equation, including spin and antiparticles.

Translate some special relativistic quantum mechanical solutions from the language of quantum wave functions to the language of the modified spacetime metric. The most elementary observations is that each plane wave eigenvalue of 4-momentum has nonmetricity $nm = 0$ and the conformal factor ρ is constant. Interpret the key features of each solution in terms of concepts of conformal geometry, instead of exclusively in terms of operators on Hilbert spaces of wave functions.

Extend the special-relativistic solutions to general relativity where metric **g** includes gravitation.

Extend the construction to concepts from quantum field theory: discreteness of particle number, and creation and annihilation operators.

## 6    Conclusions

- Quantum terms in relativistic quantum wave equations have the structure of scalar curvature of conformal dilation in the connection of spacetime, plus a divergence term that can be ignored in action integrals. Planck's reduced constant ℏ appears as a scaling factor in the conformal dilation and nowhere else in the foundations of the theory. The complex number *i* does not appear in the Lagrangian, or the wave equations. Exclusion of *i* facilitates extension of the theory to classical geometry and general relativity.
- The conformal dilation can be integrated to define a conformal factor ρ that defines a modified spacetime metric **g'** = exp(ρ) **g**. The theory can be expressed with **g'** without conformal dilation. The metrics **g'** and **g** are identical outside the domain of quantum fields. The connection preserves **g'** in classical physics and in the domain of quantum waves.
- Conformal dilation and the modified spacetime metric can be interpreted geometrically as the continuum limit of a distribution of inclusions and voids in a lattice. This interpretation fits with analogous interpretations of all other gauge fields of fundamental physics: gravitation (disclinations), intrinsic angular momentum (dislocations), and the unitary gauge theories, including electrodynamics (dispirations).
- Every field component of every free quantum field theory satisfies the KG equation, so the results can be extended to all quantum field theories. Each unique quantum field has its own amplitude, hence its own conformal dilation and conformal factor ρ.
- The metric **g'** straddles the fields of quantum theory and affine geometry. Such a hybrid construction may help to bridge the gap between gravitational theory and quantum theory.

The circle bundle structure of the symplectic model of complex phase, plus the uncertainty principle (which has a strong claim to be called the principle of least action), motivate a conjecture that any measurable action is an integer multiple of ℏ/2. Such a result might be called the "principle of discrete action."

Although this work does not address measurement processes and concepts of quantum field theory, it establishes that conformal structure arises at the level of quantum wave mechanics.

A description of quantum wave mechanics with the modified spacetime metric may help to resolve confusion about the foundations of quantum theory. A leading contemporary textbook on quantum mechanics states:

> "There is no general consensus as to what its fundamental principles are, how it should be taught, or what it really 'means'… the purpose of this book is to teach you how to *do* quantum mechanics."[7]

## 7    Acknowledgement

I would like to acknowledge advice on scope of the work and exposition from a retired physicist who wishes to remain anonymous.

**Appendix A: Computer algebra script for conformal dilation of the Klein-Gordon Lagrangian**

This executable computer algebra script provides traceable derivations of formulas in the main article.

Computer algebra system Macsyma 2.4.1a was used to verify most of the computations in this work.[8] Macsyma 2.4.1a has capabilities for variational calculus and tensor analysis that are not described in the 1996 reference manual.

# Conformal Dilations and the Klein-Gordon Equation

### Notation and terminology

The term "complex polar form" means the wave function is expressed using complex numbers in polar form. Spacetime coordinates are always Minkowski coordinates.

- * Dim = spacetime dimension >= 2, with coordinates (ct, $x^1$, ..., $x^{dim-1}$)
- * $g_{\mu\nu}$ = metric, with signature is ( + , - , - , - ). We write $g_{\mu\nu}$ and $g^{\mu\nu}$ explicitly to avoid sign ambiguities.
- * hb = Plancks reduced constant h_bar = $h/(2\pi)$
- * $\psi$ = R exp( - i S) = scalar relativistic KG field, with complex conjugate $\psi^\dagger$ (denoted $\chi$ in code)
- * m = mass $c^2$
- * p = Momentum operator = i hb d/dx
- * nm = nonmetricity covector of conformal dilations.
- * In Macsyma input code, the symbol "@" before a tensor index indicates it is an upper (contravariant) index.

Specify that g is the metric. This loads Macsyma's inidical tensor package.
Define nm as an alias for name inm that is card coded in Macsyma.
Specify that these variables are non-negative.

**(c1)**    (loadprint : false, imetric(g), alias(nm, inm), assume(\r >= 0, m >= 0, hb >= 0, \kgrav >= 0))

**(d1)**    $[R \geq 0, m \geq 0, hb \geq 0, Kgrav \geq 0]$

## 1. Klein-Gordon Lagrangian and variations in terms of $\psi$

Write the Lagrangian of the wave field in tensor index notation.
The Lagrangian has these overall factors:
- * divided by a factor of m so that the momentum tensor has units of energy-momentum.
- * Omit factor of 1/2 in Wald, so that momentum tensor contains term $p^i\, p^j\, /m$ (not $1/2\, p^i\, p^j\, /m$).

See equation (E.1.6), page 451, *General Relativity*, 1984, by R. M. Wald. Adjust for g ~ (-,+,+,+) in Wald, p 23.

**(c2)**
```
(remcomps(psi),  remcomps(chi),
 kg_lagrangian1 : (- hb^2*chi([],a)*psi([],b)*g([@a,@b]) + m^2*chi([])*psi([]))/m,
 ishow('kg_lagrangian1 = kg_lagrangian1) )
```

**(d2)**
$$kg\_lagrangian1 = \frac{\chi\, \psi\, m^2 - \chi_{,a}\, \psi_{,b}\, g^{a\,b}\, hb^2}{m}$$

Vary the Klein-Gordon action by the wave function $\psi$ and $\psi^\dagger$ (denoted $\chi$ in code).

Flush derivatives of metric $g$ to get the equation in special relativity with Minkowski coordinates.
(We will declare $g$ constant after computing the momentum tensor as the variation of the action by $g$.)

See equation (4.2.19) page 63, *General Relativity*, 1984, by R. M. Wald. Adjust for $g \sim (-,+,+,+)$ in Wald.

(c3)
```
block([ivariation_deriv_degree : 1], kg_equation1 : ivariation(kg_lagrangian1, chi([])),
  kg_equation1 : ratsimp(icanform( icontract(flushd(kg_equation1, g)))),
  kg_equation1 = 0, ishow('kg_equation1= %%) )
```

(d3)
$$\text{kg\_equation1} = \left( \frac{\psi\, m^2 + \psi_{,\%1\,\%2}\, g^{\%1\,\%2}\, \hbar^2}{m} = 0 \right)$$

The momentum tensor is the variation of the action by metric $g$, divided by $\sqrt{|\det(g)|}$.
First vary the action by $g$ to get the momentum tensor, then restrict $g$ to be the Minkowski metric.
The action integral is $\text{integrate}(L\, \sqrt{|\det(g)|})\, d\mathbf{x}$.

(c4)
```
(kg_mom0 : facsum(ivariation( kg_lagrangian1*sqrt(abs(det(g))), g([i,j]))/sqrt(abs(det(g))), hb),
  ishow('kg_mom0 = kg_mom0))
```

(d4)
$$\text{kg\_mom0} = \frac{\chi\, \psi\, g^{j\,i}\, m^2 - \chi_{,\%1}\, \psi_{,\%2}\, \hbar^2 \left( g^{\%1\,\%2}\, g^{j\,i} - g^{\%1\,i}\, g^{\%2\,j} - g^{\%1\,j}\, g^{\%2\,i} \right)}{2\, m}$$

Macsyma needs help to combine the last two terms above.

See equation (4.2.20), page 63 [*General Relativity*, 1984, by R. M. Wald]. Adjust for $g \sim (-,+,+,+)$ in Wald.

This work divides Wald's value by mass so that the momentum tensor has the units of energy-momentum.

(c5)
```
block([tmp], tmp: expand(kg_mom0), substpart( part(tmp, 3), tmp, 4),
  kg_mom1 : multthru(facsum(%%, g([@j, @i]))),
  ishow('kg_mom1 = kg_mom1))
```

(d5)
$$\text{kg\_mom1} = \frac{g^{j\,i}\left(\chi\, \psi\, m^2 - \chi_{,\%1}\, \psi_{,\%2}\, g^{\%1\,\%2}\, \hbar^2\right)}{2\, m} + \frac{\chi_{,\%1}\, \psi_{,\%2}\, g^{\%1\,i}\, g^{\%2\,j}\, \hbar^2}{m}$$

## 2. Lagrangian and its variational derivatives in complex polar form

### 2.1 Wave function and Lagrangian in complex polar form

Define the wave function $\psi$ as $R \exp(-i S)$, where $R$ = amplitude, $S$ = phase.

(c6)
```
(remcomps(psi), components(psi([]),\r([])*exp( - %i*\s([]))),
  remcomps(chi), components(chi([]), \r([])*exp(%i*\s([]))),
  ishow(['psi = psi([]), 'chi = chi([])]) )
```

(d6)
$$\left[ \psi = R\, e^{-i\,S},\ \chi = R\, e^{i\,S} \right]$$

Define pattern match rules that replace $d(S)$ with $p/\hbar$ (momentum) and replace second derivatives of $S$.
**Matchdeclare** tells the pattern matcher that a variable must be atomic to be used in a match.

(c7)
```
(matchdeclare([i%,j%],atom),
  [defrule(\s_to_p1, \s([],i%), p([i%])/hb), defrule(\s_to_p2, \s([],i%,j%), p([i%],j%)/hb)],
  ishow(%%) )
```

(d7)
$$\left[ \text{S\_to\_p1} : S_{,i\%} \to \frac{p_{i\%}}{\hbar},\ \text{S\_to\_p2} : S_{,i\%\,j\%} \to \frac{p_{i\%,j\%}}{\hbar} \right]$$

Express Klein-Gordon Lagrangian in terms of amplitude R and momentum **p**.

Apply the pattern matching rule to the Lagrangian to replace d(S) with - **p**/hb.

The structure of the Lagrangian is $R^2 (m^2 - p^2) - R^2 *$ (quantum terms).

(c8) (kg_lagr_polar1 : icanform(expand(apply1(eval(kg_lagrangian1), \s_to_p1))),
    ishow('kg_lagr_polar1 = kg_lagr_polar1) )

(d8)
$$kg\_lagr\_polar1 = R^2 m^2 - \frac{R_{,\%1} R_{,\%2} g^{\%1 \%2} hb^2}{m} - \frac{R^2 p_{\%1} g^{\%1 \%2} p_{\%2}}{m}$$

## 2.2 Wave equation in terms of amplitude and momentum vector

Multiply the wave equation by the conjugate wave function and express result in terms of R and **p**.

(c9) (kg_eqn_polar1 : rectform(facsum(flushd(apply1(eval(chi([])*kg_equation1),\s_to_p2,\s_to_p1), g),hb)),
    ishow('kg_eqn_polar1 = kg_eqn_polar1) )

(d9)
$$kg\_eqn\_polar1 = \frac{R^2 \left(m^2 - p_{\%1} g^{\%1 \%2} p_{\%2}\right) + R R_{,\%1 \%2} g^{\%1 \%2} hb^2}{m}$$
$$- \frac{i R g^{\%1 \%2} \left(R_{,\%1} p_{\%2} + R p_{\%1,\%2} + R_{,\%2} p_{\%1}\right) hb}{m}$$

### 2.2.1 Real part of the wave equation

Real part of KG equation is the off-shell quantum version of the relativistic momentum constraint $p^2 - m^2 = 0$.

Since g ~ (+, -, -, -), the quantum term $hb^2 R R_{j,k} g^{jk}$ is positive near the center of wave packets.

This means that the extra energy content of a wave packet is positive.

(c10) (kg_eqn_polar_real1 : realpart(kg_eqn_polar1),
    ishow('kg_eqn_polar_real1 = kg_eqn_polar_real1) )

(d10)
$$kg\_eqn\_polar\_real1 = \frac{R^2 \left(m^2 - p_{\%1} g^{\%1 \%2} p_{\%2}\right) + R R_{,\%1 \%2} g^{\%1 \%2} hb^2}{m}$$

### 2.2.2 Imaginary part of the wave equation

Imaginary part of kg_eqn_polar1 is quantum version of the momentum conservation law $div(r^2 p) = 0$.

(c11) (kg_eqn_polar_imag1 : ratsimp(icanform(imagpart(kg_eqn_polar1)), m, hb),
    ishow('kg_eqn_polar_imag1 = kg_eqn_polar_imag1) )

(d11)
$$kg\_eqn\_polar\_imag1 = - \frac{\left(2 R R_{,\%1} g^{\%1 \%2} p_{\%2} + R^2 p_{\%1,\%2} g^{\%1 \%2}\right) hb}{m}$$

## 2.5 Momentum tensor in terms of amplitude and momentum vector

The momentum tensor is vary(action, metric).

kg_mom1 expresses the momentum tensor in terms of wave function $\psi$ and its conjugate.

kg_mom_polar1 expresses the momentum tensor in complex polar form.

(c12)
```
(kg_mom_polar0 : facsum(icanform(expand(ev(kg_mom1)))),
   ishow( 'kg_mom_polar0 = kg_mom_polar0) )
```

(d12)
$$kg\_mom\_polar0 = \frac{\begin{pmatrix} R^2 g^{ij} m^2 - R^2 S_{,\%1} S_{,\%2} g^{\%1 \%2} hb^2 g^{ij} - R_{,\%1} R_{,\%2} g^{\%1 \%2} hb^2 \\ * g^{ij} + 2 R^2 S_{,\%1} S_{,\%2} g^{\%1 i} g^{\%2 j} hb^2 - 2i R R_{,\%1} S_{,\%2} g^{\%1 i} g^{\%2 j} \\ * hb^2 + 2 R_{,\%1} R_{,\%2} g^{\%1 i} g^{\%2 j} hb^2 + 2i R R_{,\%1} S_{,\%2} g^{\%1 j} g^{\%2 i} hb^2 \end{pmatrix}}{2 m}$$

Below is the imaginary part of kg_mom_polar0, which equals zero.

Macsyma utlities for applying tensor symmetries do not know how to simplify these terms.

* The first term = 0. Macsyma is unable to determine that $(u^j v^k - v^j u^k) g_{jk} = 0$.

* The second term = 0, because it results from variation of action by $g_{ij}$, which Macsyma knows is symmetric.

(c13)  ishow(trigsimp(expand(demoivre(imagpart(ev(kg_mom0))))))

(d13)
$$\frac{\begin{pmatrix} (R R_{,\%1} S_{,\%2} - R S_{,\%1} R_{,\%2}) g^{\%1 \%2} hb^2 g^{ji} \\ + \left((R S_{,\%1} R_{,\%2} - R R_{,\%1} S_{,\%2}) g^{\%1 i} g^{\%2 j} + (R S_{,\%1} R_{,\%2} - R R_{,\%1} S_{,\%2}) g^{\%1 j} g^{\%2 i}\right) hb^2 \end{pmatrix}}{2 m}$$

What remains of kg_mom_polar0 is the real part, which is manifestly symmetric.

(c14)
```
(kg_mom_polar1 : trigsimp(expand(demoivre(realpart(ev(kg_mom0))))),
   ishow('kg_mom_polar1 = kg_mom_polar1))
```

(d14)
$$kg\_mom\_polar1$$
$$= \frac{\begin{pmatrix} R^2 g^{ji} m^2 + \left(-R^2 S_{,\%1} S_{,\%2} - R_{,\%1} R_{,\%2}\right) g^{\%1 \%2} hb^2 g^{ji} \\ + \left((R^2 S_{,\%1} S_{,\%2} + R_{,\%1} R_{,\%2}) g^{\%1 i} g^{\%2 j} + (R^2 S_{,\%1} S_{,\%2} + R_{,\%1} R_{,\%2}) g^{\%1 j} g^{\%2 i}\right) hb^2 \end{pmatrix}}{2 m}$$

kg_mom_polar2 expresses the momentum tensor in terms of amplitude R and momentum vector p.

(c15)
```
(icanform(icontract(expand(apply1(kg_mom_polar1, \s_to_p1)))),
   kg_mom_polar2 : multthru(facsum(%%, hb)),
   ishow('kg_mom_polar2 = kg_mom_polar2) )
```

(d15)
$$kg\_mom\_polar2 = \frac{R^2 \left(g^{ij} m^2 + 2 p^i p^j - p_{\%1} p^{\%1} g^{ij}\right)}{2 m}$$
$$- \frac{R_{,\%1} R_{,\%2} hb^2 \left(g^{\%1 \%2} g^{ij} - g^{\%1 i} g^{\%2 j} - g^{\%1 j} g^{\%2 i}\right)}{2 m}$$

The last two terms in the numerator are identical. We help Macsyma to make this simplification.

**(c16)** (substpart(g([@%1, @i])*g([@%2, @j]), kg_mom_polar2, 2, 1,1, 4, 3,1),
kg_mom_polar3 : %%, ishow('kg_mom_polar3 = kg_mom_polar3) )

**(d16)**
$$\text{kg\_mom\_polar3} = \frac{R^2 \left( g^{ij} m^2 + 2 p^i p^j - p_{\%1} p^{\%1} g^{ij} \right)}{2m} - \frac{R_{,\%1} R_{,\%2} hb^2 \left( g^{\%1 \%2} g^{ij} - 2 g^{\%1 i} g^{\%2 j} \right)}{2m}$$

## 3. Lagrangian from conformal dilations

Include conformal dilations in the geometry. The switch **inonmet_flag** turn on this feature.

The dilation covector is denoted **nm** .

Specify that metric **g** is constant, so the computations are for special relativity, without gravitation.

Variation with respect to **g** will not work when **g** is declared constant.

**(c17)**    (inonmet_flag : true, declare(g, constant))$

### 3.1 Covariant derivatives with conformal dilations

This subsection contains background information not direclty needed in the computations. It is included to help familiarize the reader with conformal dilataion in an otherwise flat space.

Covariant derivative of metric **g** in the $x^k$ direction, expressed with connection coefficients.

*   **inmc**$_{kj}{}^i$ are the the contributions of the conformal dilations to the connection coefficients.
*   **icchr2** are the indicial Christoffel connection coefficients. Since **g** is constant, the **icchr2** vanish.

**(c18)**    (rename(eval(covdiff(g([i,j]), k))), ishow('covdiff(g([i,j]),k) = %%))

**(d18)**
$$\text{covdiff}\left(g_{ij}, k\right) = -g_{i\,\%1}\left(\text{inmc2}_{kj}{}^{\%1} + \text{ichr2}_{kj}{}^{\%1}\right) - g_{\%1\,j}\left(\text{inmc2}_{ki}{}^{\%1} + \text{ichr2}_{ki}{}^{\%1}\right)$$

Covariant derivative of metric **g** in the $x^k$ direction, expressed with the conformal dilation covector $\text{nm}_k$.

This shows that $\text{nm}_k$ is the expansion factor when **g** is parallel translated in direction $d/dx^k$.

**(c19)**    (icanform(icontract(expand(rename(infeval(covdiff(g([i,j]), k))))))),
ishow('covdiff(g([i,j]),k)  = %%))

**(d19)**
$$\text{covdiff}\left(g_{ij}, k\right) = -g_{ij}\,\text{nm}_k$$

Covariant derivative of vectorfield **v** in the $x^k$ direction, expressed with connection coefficients:

**(c20)**    (icanform(icontract(expand(rename(eval(covdiff(v([@j]), k))))))),
ishow('covdiff(v([@j]), k) = %%))

**(d20)**
$$\text{covdiff}\left(v^j, k\right) = v^{\%1}\,\text{inmc2}_{k\,\%1}{}^j + v^j{}_{,k} + \text{ichr2}_{\%1\,k}{}^j v^{\%1}$$

Covariant derivative of vectorfield **v** in the $x^k$ direction, expressed with the conformal dilation covectorfield $\text{nm}_k$:

**(c21)**    (icanform(icontract(expand(rename(infeval(covdiff(v([@j]), k))))))),
ishow('covdiff(v([@j]), k)  = %%))

**(d21)**
$$\text{covdiff}\left(v^j, k\right) = \frac{\text{nm}_{\%1}\,v^{\%1}\,\text{kdelta}_k{}^j}{2} - \frac{\text{nm}_{\%1}\,g^{\%1\,j}\,v_k}{2} + \frac{v^j\,\text{nm}_k}{2} + v^j{}_{,k}$$

Covariant derivative of covectorfield ω in the $x^k$ direction, expressed with connection coefficients:

**(c22)** (rename(eval(covdiff(omega([j]), k))), ishow('covdiff(omega([j]), k) = %%))

**(d22)**
$$\mathrm{covdiff}\left(\omega_j, k\right) = \omega_{j,k} - \omega_{\%1}\left(\mathrm{inmc2}_{kj}{}^{\%1} + \mathrm{ichr2}_{kj}{}^{\%1}\right)$$

Covariant derivative of covectorfield ω in the $x^k$ direction, expressed with the conformal dilation vectorfield $nm_k$:

**(c23)** (icanform(icontract(expand(rename(infeval(covdiff(omega([j]), k))))))), ishow('covdiff(omega([j]), k) = %%))

**(d23)**
$$\mathrm{covdiff}\left(\omega_j, k\right) = -\frac{nm_j\,\omega_k}{2} - \frac{\omega_j\,nm_k}{2} + \frac{nm_{\%1}\,\omega^{\%1}\,g_{jk}}{2} + \omega_{j,k}$$

### 3.2 Scalar curvature of conformal dilations

Start with the scalar curvature function **iscurvature** in Macsyma. It needs simplifcation so we do not display it.

**(c24)** (scalar_curv1 : rename(eval(iscurvature([]))), ishow('scalar_curv1 = scalar_curv1))

**(d24)**
$$\mathrm{scalar\_curv1} = g^{\%1\,\%3}\,\mathrm{icurvature}_{\%1\,\%2\,\%3}{}^{\%2}$$

Start with the scalar curvature function **iscurvature** in Macsyma, and perform these simplifications.

a) **infeval** converts connection coeffiicents to nonmetricity covector **nm**.
b) **icanform** puts indices in standard order: example: g([i,j]) and g([j,i]) both become g([i, j]).
c) **icontract** contracts tensor indices.
d) **eval** makes some terms that are identical combine or cancel.
e) **rename** standardizes contracted index names. Example: nm[a]*g[@a, @b]*nm[b] and nm[c]*g[@c, @d]*nm[d] both become nm[%1]*g[@%1, @%2]*nm[%2].
f) **icanform** restores identical represention of some identical terms.
g) **facsum** orgainzies factors and sums to simplfy the final expression.

**(c25)** (multthru(facsum(icanform(rename(eval(icontract(icanform(infeval(iscurvature([]))))))), nm([%1]))), scalar_curv2 : %%, ishow('scalar_curv2 = scalar_curv2))

**(d25)**
$$\mathrm{scalar\_curv2} = -\frac{nm_{\%1}\,g^{\%1\,\%2}\,nm_{\%2}\,(\dim - 2)\,(\dim - 1)}{4} - nm_{\%1,\%2}\,g^{\%1\,\%2}\,(\dim - 1)$$

Special relativistic scalar curvature with conformal dilation for four spacetime dimensions.

**(c26)** (scalar_curv2_4d : expand(subst(dim=4, scalar_curv2)), ishow(scalar_curv2_4d))

**(d26)**
$$-\frac{3\,nm_{\%1}\,g^{\%1\,\%2}\,nm_{\%2}}{2} - 3\,nm_{\%1,\%2}\,g^{\%1\,\%2}$$

### 3.3 Conformal dilation covector for Klein-Gordon field

The essence of the definition of the conformal dilation covector nm is the product of the gradient of amplitude R of the wave function and Planck's reduced constant hb = h/(2 π).

The field equations of general relativity are $G_{ab}$ / (2 Kgrav) = (momentum$_{ab}$), where Kgrav = (8 π G / $c^4$). nm includes a factor sqrt(2 Kgrav) to cancel this out, so nm fits into general relativity with correct units. The factors in dim enable nm to replicate the quantum term in the KG equation in spacetime of any dimension.

(c27) (remcomps(nm), components(nm([k]), 2*sqrt(2*\kgrav)/sqrt((m*(dim-2)*(dim-1)))* hb* \r([], k)),
     block([tmp], tmp([k]), subst('nm, tmp, %%)), ishow(%% = nm([k])) )

(d27)
$$nm_k = \frac{2\sqrt{2}\, R_{,k}\, hb\, \sqrt{Kgrav}}{\sqrt{(dim-2)(dim-1)}\, \sqrt{m}}$$

### 3.4 Scalar curvature with dilation covector of Klein-Gordon field

Compute the scalar curvature using the assigned value of nm.

(c28) (scalar_curv3 : facsum((expand(eval(scalar_curv2))), \k[grav], hb),
     ishow('scalar_curv3 = scalar_curv3))

(d28)
$$scalar\_curv3 = \frac{\left(-2 R_{,\%1} R_{,\%2} g^{\%1\,\%2} (dim-2)\, hb^2\, Kgrav - 2\sqrt{2}\, *\, R_{,\%1\,\%2}\, g^{\%1\,\%2}\, \sqrt{dim^2 - 3\, dim + 2}\, hb\, \sqrt{m}\, *\sqrt{Kgrav}\right)}{(dim-2)\, m}$$

The Laplacian of a function is a divergence so it contributes only a boundary term in the action integral. So it can be omitted from the Lagrangian without affecting the action integral or its variational derivatives such as the field equation or the momentum tensor.

(c29) (factor(expand(ratsubst(0, \r([],%1,%2), rncombine(expand(eval(scalar_curv3)))))),
     scalar_curv4 : %%, ishow('scalar_curv4 = scalar_curv4))

(d29)
$$scalar\_curv4 = -\frac{2 R_{,\%1} R_{,\%2}\, g^{\%1\,\%2}\, hb^2\, Kgrav}{m}$$

Compute scalar curvature divided by 2 Kgrav for a spacetime of any dimension dim.

(c30) ( scalar_curv4_div2k : scalar_curv4/(2*\kgrav),
     ishow('scalar_curv4/(2*\kgrav) = scalar_curv4_div2k))

(d30)
$$\frac{scalar\_curv4}{2\, Kgrav} = -\frac{R_{,\%1} R_{,\%2}\, g^{\%1\,\%2}\, hb^2}{m}$$

## Appendix B: Affine connections with metric and conformal dilation

### B.1    Riemannian metric connection

A metric affine connection can be uniquely defined by specifying a metric tensor.
The Levi-Civita connection (L.C. connection) of g has connection coefficients called Christoffel symbols of the second kind

(B-1)    $^{LC}\Gamma_{\mu\beta}{}^{\alpha} = \tfrac{1}{2} g^{\alpha\gamma} (g_{\gamma\beta,\mu} + g_{\mu\gamma,\beta} - g_{\mu\beta,\gamma})$ .

### B.2    Riemann–Cartan connection with specified metric and torsion

A Riemann–Cartan connection has a metric and affine torsion Tr that preserves the metric, with connection coefficients

(B-2) $$^{RC}\Gamma_{\mu\beta}{}^\alpha = {}^{LC}\Gamma_{\mu\beta}{}^\alpha - KT_{\mu\beta}{}^\alpha$$

where the contortion tensor KT is given in terms of the affine torsion $Tr_{\mu\nu}{}^\alpha$ by

(B-3) $$KT_{\mu\beta}{}^\alpha = -\tfrac{1}{2}(Tr_{\mu\beta}{}^\alpha + Tr^\alpha{}_{\mu\beta} + Tr^\alpha{}_{\beta\mu}).$$

Torsion Tr is written in any of three equivalent forms. Torsion itself is the translational curvature in affine geometry, and is the most fundamental of the three tensors. The contortion tensor KT is the contribution of torsion to the connection coefficients. The "modified torsion" equals intrinsic angular momentum in Einstein–Cartan theory.

Torsion is needed in spacetime physics only when fields with intrinsic angular momentum are present. We will not include torsion elsewhere in this work.

**B.3    Affine connection with specified metric, torsion and conformal dilation**

The general nonmetricity tensor CW is a rank 3 covariant tensor field which is symmetric in its last two indices. It is otherwise unrestricted. Some basic computational results involving nonmetricity are:

(B-4) $$g_{\mu\alpha;\beta} = g_{\mu\alpha|\beta} - g_{\lambda\alpha} CW_{\beta\mu}{}^\lambda - g_{\mu\lambda} CW_{\beta\alpha}{}^\lambda$$

where a vertical bar ( | ) denotes covariant differentiation with respect to the metric connection.

The full connection coefficients cc with Riemannian metric, torsion and conformal nonmetricity are:

(B-5) $$cc_{\mu\beta}{}^\alpha = {}^{LC}\Gamma_{\mu\beta}{}^\alpha - KT_{\mu\beta}{}^\alpha + CW_{\mu\beta}{}^\alpha$$

Conformal dilation is a special case of nonmetricity in which the covariant derivative of the metric g is proportional to g.

(B-6) $$g_{\mu\alpha;\beta} = -CW_{\beta\mu\alpha} = -nm_\beta\, g_{\mu\alpha}$$

A manifold with metric g and conformal dilation nm is called a Weyl geometry.

Conformal dilations are a special case of conformal transformations The vector field $nm_\beta$ is the conformal dilation covector field. This nonmetricity is called conformal dilation because angles are preserved by parallel translation, although lengths are not. In this case, nonmetricity term $CW_{\mu\beta}{}^\alpha$ in the connection form reduces to the conformal dilation

(B-7) $$CW_{\mu\beta}{}^\alpha = \tfrac{1}{2}(nm_\mu\, \delta^\alpha{}_\beta + nm_\beta\, \delta_\mu{}^\alpha - nm\, g_{\mu\beta}).$$